\documentclass[preprint,preprintnumbers,nofootinbib,aps,10pt,twocolumn]{revtex4-1}
\usepackage{amsmath,amssymb,bm,epsfig}
\usepackage{color}
\usepackage{natbib}
\usepackage{hyperref} 
\usepackage{ulem}
\usepackage{graphicx}
\usepackage{xcolor,colortbl}
\RequirePackage{lineno}
\newcommand{\nn}{\nonumber}
\newcommand{\sNN}{\sqrt{s_{\textrm{NN}}}}

\newcommand{\ka}{\rm{K}}

\newcommand{\prp}{\textrm{p}}

\definecolor{Gray}{gray}{0.85}
\newcolumntype{a}{>{\columncolor{Gray}}c}

\def \beq{\begin{equation}}
\def \eeq{\end{equation}}
\def \beqa{\begin{eqnarray}}
\def \eeqa{\end{eqnarray}}

\def \dv1{$\Delta v_1$}

\begin{document}

\title{Baryon inhomogeneities driven charge dependent directed flow in heavy ion collisions}

\author{Tribhuban Parida}
\email{tribhubanp18@iiserbpr.ac.in}
\author{Sandeep Chatterjee}
\email{sandeep@iiserbpr.ac.in}

\affiliation{Department of Physical Sciences,\\
Indian Institute of Science Education and Research Berhampur,\\ 
Transit Campus (Govt ITI), Berhampur-760010, Odisha, India}

\begin{abstract}
Electromagnetic field in heavy ion collisions are expected to cause charge dependent directed flow 
splitting (\dv1). Such charge dependent \dv1 has been observed by the STAR collaboration. We 
demonstrate that relativistic dissipative fluid dynamic simulations with baryon diffusion that include realistic 
model of baryon stopping in the initial condition and no contribution from electromagnetic field describe the measured \dv1 
for observables involving baryons and anti-baryons. This suggests strong background contribution from 
baryon current as a response to initial state baryon inhomogeneities to such charge dependent \dv1 
involving baryons and anti-baryons. Our current model calculations that only account for the evolution 
of the baryon charge and not electric charge and strangeness miss the observed \dv1 of mesons 
leaving their interpretation open.
\end{abstract}

\maketitle

\section{Introduction}

An intense electromagnetic (EM) field with a strength of approximately $10^{18}$ G is likely to be generated in non-central heavy ion collisions~\cite{Deng:2012pc,McLerran:2013hla, Voronyuk:2011jd, Kharzeev:2007jp, Skokov:2009qp, Tuchin:2013ie}. The primary source of the EM field is the motion of the oppositely moving charged spectators, although the participants also contribute significantly~\cite{Kharzeev:2007jp, Gursoy:2018yai}. Detecting the signals of EM field is challenging as the field decays rapidly over time~\cite{Tuchin:2013apa}. Efforts to detect the signals of EM field and 
explore the associated rich phenomenology is still on~\cite{STAR:2009wot,ALICE:2020siw,Burnier:2011bf,Kharzeev:2010gd,Fukushima:2008xe,Zhao:2019hta,Li:2020dwr,STAR:2021mii,Bzdak:2019pkr}. One such phenomena is the flow separation of oppositely charged hadrons~\cite{Hirono:2012rt,Nakamura:2022ssn,Das:2016cwd,STAR:2019clv,ALICE:2019sgg,STAR:2016cio,Bzdak:2019pkr}.

The charged constituents formed inside the fireball after the collision are expected to be affected by the electric fields which are produced due to three distinct effects :  
Coulomb, Farady and Lorentz~\cite{Gursoy:2018yai}. 
A non-zero electric field is produced inside the medium due to the spectators, which exert a Coulomb force on the charged 
objects. Secondly, as the magnetic field produced during the collision decays with time, it creates another component of electric field in the reaction plane due to Faraday induction.
Additionally, the medium expands with a strong longitudinal flow, and when boosted to the local rest frame of the fluid, the lab frame magnetic field produces an electric field in the fluid rest frame. The force acting on the charged constituents due to this electric field is equivalent to the Lorentz force in lab frame. All three electric fields produced due to these effects are parallel to the impact parameter which we label as the x-axis. Hence, the force generated due to the resultant field pushes opposite charges in opposite directions of x-axis. This flow separation may be reflected in the final momentum distribution of the produced particles and can be quantified through the measurement of the splitting of directed flow ($\Delta v_1$) between oppositely charged hadrons. The directed flow ($v_1$) is defined as the first harmonic
coefficient in the Fourier series expansion of the azimuthal
distribution of the particles produced relative to the reaction plane, $\Psi_{RP}$.
\beq
\frac{d^2 N}{dy d\phi} = \frac{dN}{dy}  \left( 1 + 2 v_1(y) cos ( \phi - \Psi_{RP} ) + ... \right)
\eeq
where $\phi$ and $y$ are the azimuthal angle and rapidity of the produced particle respectively.

Recently STAR collaboration has observed the splitting of $v_1$
between  $\pi^{+}-\pi^{-}$, $K^{+}-K^{-}$ and $p-\bar{p}$ and observed a non trivial dependency along centrality~\cite{STAR:2023jdd}. 
This has been interpreted as a consequence of the initial EM field, where the interplay between Coulomb, Faraday and Lorentz effect shows the non-trivial dependency of $\Delta v_1$ along centrality. However, this kind of splitting may not be solely due to EM field. There exists another crucial physical phenomena which can cause such splitting between the aforementioned pairs. We are describing that phenomena below.  

It is important to note that in relativistic heavy ion collisions, a substantial amount of net-baryon and electric charge is deposited inside the fireball, along with energy~\cite{STAR:2017sal,BRAHMS:2003wwg,BRAHMS:2009wlg,NA49:2010lhg}.
However, the distribution of these conserved charges within the fireball differs significantly depending on the collision energies. Specifically, more baryons are stopped in the mid-rapidity region for low-energy collisons as compared to high-energy ones, which is referred to as baryon stopping in nuclear collisions~\cite{Ranft:2000sf,Mehtar-Tani:2009wji,Capella:1999cz,Li:2016wzh,Li:2018ini,Bialas:2016epd}. 

From various model calculations it has been shown that, initial inhomogeneous distribution of net-baryon leads to a $v_1$ splitting between proton and anti-proton~\cite
{Bozek:2022svy,Du:2022yok,Parida:2022zse,Parida:2022ppj,Jiang:2023fad}.
Similarly, the inhomogeneous distribution and subsequent evolution of net strangeness and electric charge can produce  
splitting between $K^{+}-K^{-}$ or $\pi^{+}-\pi^{-}$. 

In realistic non-central heavy-ion collision events, both the effect of conserved charge stopping and the EM field contribute to the $v_1$ splitting between $\pi^{+}-\pi^{-}$, $K^{+}-K^{-}$ and $p-\bar{p}$~\cite{Gursoy:2018yai,Parida:2022zse,Parida:2022ppj,Nakamura:2022ssn}. To observe the effect of the EM field, one must remove the background contribution from the other source. In this work, we have shown the contribution of conserved charge stopping to $\Delta v_1$ is substantial. Therefore, a more detailed investigation is required to accurately quantify the EM field signals in the presence of the background of conserved charge stopping.

The model used in our study provides a consistent description of the $v_1$ of identified hadrons which is shown in our previous studies~\cite{Parida:2022zse,Parida:2022ppj}. Especially, it captures
the splitting of $v_1$ between proton and anti-proton well. However, the model has certain limitations, as it only considers the independent evolution of net-baryon density, while strangeness and electric charge evolution are constrained by the baryon density. Nevertheless, our findings provide compelling evidence that conserved charge stopping substantially impacts $\Delta v_1$, which complicates the extraction of EM field signals from this observable.      

\section{Framework}

In this section, we present the multi-stage hybrid framework used in this study. We begin by describing the initial condition model, which is an input for hydrodynamic evolution. We also discuss the various model parameters.

\subsection{Initial condition}
In this study, an event-averaged smooth profile was prepared in the transverse plane of both participant and binary collision densities by averaging over the participant and binary collision sources obtained from event-by-event Monte Carlo Glauber model over 25,000 initial configurations at a given centrality class. This process allowed us to obtain an ensemble-averaged initial density distribution that represents the entire collision system~\cite{Shen:2020jwv}. At the beginning of the hydrodynamic evolution, the three-dimensional distribution of energy density ($\epsilon$) at a constant proper time ($\tau_0$) takes the following form~\cite{Bozek:2010bi}.
\beqa
  \epsilon(x,y,\eta_{s}; \tau_0) &=& \epsilon_{0} \left[ \left( N_{+}(x,y) f_{+}(\eta_{s}) + N_{-}(x,y) f_{-}(\eta_{s})  \right)\right.\nn\\
                           &&\left.\times \left( 1- \alpha \right) + N_{bin} (x,y)  \left(\eta_{s}\right) \alpha \right] \epsilon_{\eta_s}
 \label{eq.tilt}
\eeqa
Here, $N_+(x,y)$ and $N_-(x,y)$ are the participant densities of the forward and backward moving nucleus, respectively. $N_{\rm{bin}}(x,y)$ is the contribution from binary collision sources. The parameter $\alpha$ controls the relative contribution of participants and binary collision sources to the energy density. The function $f_{+,-}(\eta_s)$ introduces the asymmetric deposition of matter in the forward and backward space-time rapidity, denoted by $\eta_s$.
\begin{equation}
    f_{+}(\eta_s) = 
    \begin{cases}
    0, & \text{if } \eta_{s} < -\eta_{m}\\
    \frac{\eta_{s} + \eta_{m }}{2 \eta_{m}},  & \text{if }  -\eta_{m} \le \eta_{s} \le \eta_{m} \\
    1,& \text{if }  \eta_{m} < \eta_{s}
\end{cases}
\end{equation}
with 
\begin{equation}
    f_{-} (\eta_s) = f_{+}(-\eta_s)
\end{equation}
$\epsilon_{\eta_s}$ is a parmetrized function with two free parameters $\eta_0$ and $\sigma_{\eta}$ which is symmetric about $\eta_s=0$.
\begin{equation}
  \epsilon_{\eta_s}(\eta_s) = \exp \left(  -\frac{ \left( \vert \eta_{s} \vert - \eta_{0} \right)^2}{2 \sigma_{\eta}^2}   
    \theta (\vert \eta_{s} \vert - \eta_{0} ) \right)
    \label{eq.seven}
\end{equation}
This type of initial energy deposition leads to a tilted profile in the reaction plane~\cite{Bozek:2010bi}. The parameter $\eta_m$ is a free parameter in the model that determines the tilt of the profile. A smaller value of $\eta_m$ results in a larger asymmetric deposition of matter along the rapidity by oppositely moving participants. This leads to a larger tilt and, consequently, a larger asymmetric energy distribution along the $x$-axis in a non-zero rapidity region. Such an initial profile is successful in explaining the rapidity-dependent charged particle directed flow phenomenon~\cite{Bozek:2010bi}..

We have used our recently proposed baryon deposition ansatz~\cite{Parida:2022zse,Parida:2022ppj}. 
\begin{equation}
    n_{B} \left( x, y, \eta_s \right) = 
       N_{B} \left[ W_{+}^{B}(x,y) f_{+}^{B}(\eta_{s}) + W_{-}^{B}(x,y) f_{-}^{B}(\eta_{s})  \right]
    \label{my_baryon_ansatz}
\end{equation}
where the transverse distribution($W_{\pm}^{B}(x,y)$) of net-baryon density is not taken same as participants (initial baryon carriers) 
distribution ($N_{\pm}(x,y)$) rather we have taken a combination of both participants and binary collision sources
\begin{equation}
W_{\pm}^{B}(x,y) = \left( 1 - \omega \right) N_{\pm}(x,y) + \omega N_{coll}(x,y)
    \label{weight_ansatz_1_for_baryon}
\end{equation}
The net baryon density rapidity profiles $f_{+}^{n_{B}}$, $f_{-}^{n_{B}}$  are taken same as in~\cite{Shen:2020jwv,Denicol:2018wdp}, 
\beqa
 f_{+}^{n_{B}} \left( \eta_s \right) &=&  \left[  \theta\left( \eta_s - \eta_{0}^{n_{B} } \right)   \exp{- \frac{\left( \eta_s - \eta_{0}^{n_{B} }  \right)^2}{2 \sigma_{B, + }^2}}   + \right.\nn\\ && \left. \theta\left(  \eta_{0}^{n_{B} } - \eta_s \right)   \exp{- \frac{\left( \eta_s - \eta_{0}^{n_{B} }  \right)^2}{2 \sigma_{B, - }^2}}   \right]
\eeqa
and
\beqa
    f_{-}^{n_{B}} \left( \eta_s \right) &=&  \left[   \theta\left( \eta_s + \eta_{0}^{n_{B} } \right)   \exp{- \frac{\left( \eta_s + \eta_{0}^{n_{B} }  \right)^2}{2 \sigma_{B, - }^2}}   + \right.\nn\\ && \left. \theta\left( -\eta_s -  \eta_{0}^{n_{B} }  \right)   \exp{- \frac{\left( \eta_s + \eta_{0}^{n_{B} }  \right)^2}{2 \sigma_{B, + }^2}}   \right]
\eeqa
The normalization factor $N_B$ in Eq.~\ref{my_baryon_ansatz} is fixed by the following condition
\begin{equation}
      \int  \tau_{0}  n_{B} \left( x, y, \eta; \tau_{0} \right) dx  dy  d\eta  = \int N_{+}(x,y) + N_{-}(x,y) dx dy 
      \label{net_baryon_is_npart_2}
\end{equation}

This type of baryon distribution results in a tilted baryon profile in the reaction plane, although the tilt is different from the matter tilt. In this model, the parameter $\omega$ is a free parameter that controls the tilt of the baryon profile. In our previous work~\cite{Parida:2022zse,Parida:2022ppj}, we have shown the effect of $\omega$ on the initial baryon distribution and final-stage baryon-specific observables.

The Bjorken flow ansatz has been adopted at the starting of hydrodynamic evolution in this work. Hence, the initial fluid four velocity $u^{\mu}$ takes the following form in cartesian coordinate.
\begin{equation*}
u^{\mu} = (\cosh \eta_s,0,0,\sinh \eta_s)
\end{equation*}

\subsection{Hydrodynamic evolution and transport coefficients}
The hydrodynamic evolution of the above discussed initial profiles have been performed using the publicly available code MUSIC~\cite{Schenke:2010nt,Schenke:2011bn}.
This code solves the conservation equations for the energy-momentum tensor and net-baryon current.

The evolution equations of conserved quantities and dissipative currents that are solved in MUSIC at finite baryon densities have been described in~\cite{Denicol:2018wdp}. Strangeness and electric charge are not evolved independently in the code; instead, they are locally constrained by strangeness neutrality and a fixed electric charge to baryon density ratio. This constraint is imposed by the used EoS during the hydrodynamic evolution~\cite{Monnai:2019hkn}. 

For this study, we have set the specific shear viscosity ($C_{\eta} = \frac{ \eta T }{ \epsilon + P }$) to 0.08 in the simulation and neglected the effect due to bulk viscosity. Another crucial transport coefficient in a baryon-rich medium is the baryon diffusion coefficient ($\kappa_B$). In this work, we use a temperature ($T$) and baryon chemical potential ($\mu_B$) dependent $\kappa_B$ derived from Boltzmann equation in relaxation time approximation~\cite{Denicol:2018wdp}, as follows:
\beq
\kappa_{B} = \frac{C_B}{T} n_{B} \left[ \frac{1}{3} \coth{\left(\frac{\mu_B}{T}\right)} - \frac{n_B T}{\epsilon + p} \right]
\label{kappaB}
\eeq
Here, $C_B$ is a free parameter which we have set to unity in this study.

\subsection{Particlization and hadronic transport}
The particlization have been performed on a constant energy density ($\epsilon_f$) hypersurface using the iSS code~\cite{https://doi.org/10.48550/arxiv.1409.8164,https://github.com/chunshen1987/iSS}. We have choosen $\epsilon_f=0.26$ GeV/fm$^3$. After the primordial hadrons were sampled, they were passed through the hadronic transport code UrQMD~\cite{Bass:1998ca, Bleicher:1999xi} for further late-stage elastic or inelastic interactions.
 
\section{Results}

\begin{table}[ht]
\begin{tabular}{|p{0.6cm}|p{1.2cm}|p{0.5cm}|p{0.5cm}|p{0.5cm}|p{0.7cm}|p{0.7cm}|p{0.6cm}|p{0.6cm}|}
\hline  
$\tau_0$ \tiny{(fm)} &$\epsilon_{0}$  \tiny{(GeV/fm$^{3}$)} &$\eta_{0}$ & $\sigma_{\eta}$ &  $\eta_{0}^{n_{B}}$ & $\sigma_{B,-}$ & $\sigma_{B,+}$ & $\omega$ & $\eta_m$ \\
\hline  
 1.2 & 2.4 & 1.3 & 0.7 & 2.3 & 1.1 & 0.2 & 0.11 & 1.1 \\ 
\hline
\end{tabular}
\caption{Parameters used in the simulations. }
\label{param_table}
\end{table} 

We have studied Au+Au collisions at $\sNN = 27$ GeV. To select the model parameters, we followed the same procedure outlined in our previous work~\cite{Parida:2022zse}. The parameters are carefully chosen to capture the charged particle and net proton yield, as well as to simultaneously capture the $v_1$ of $\pi^{-}$, $p$, and $\bar{p}$.

\begin{figure}
 \begin{center}
  \includegraphics[scale=0.52]{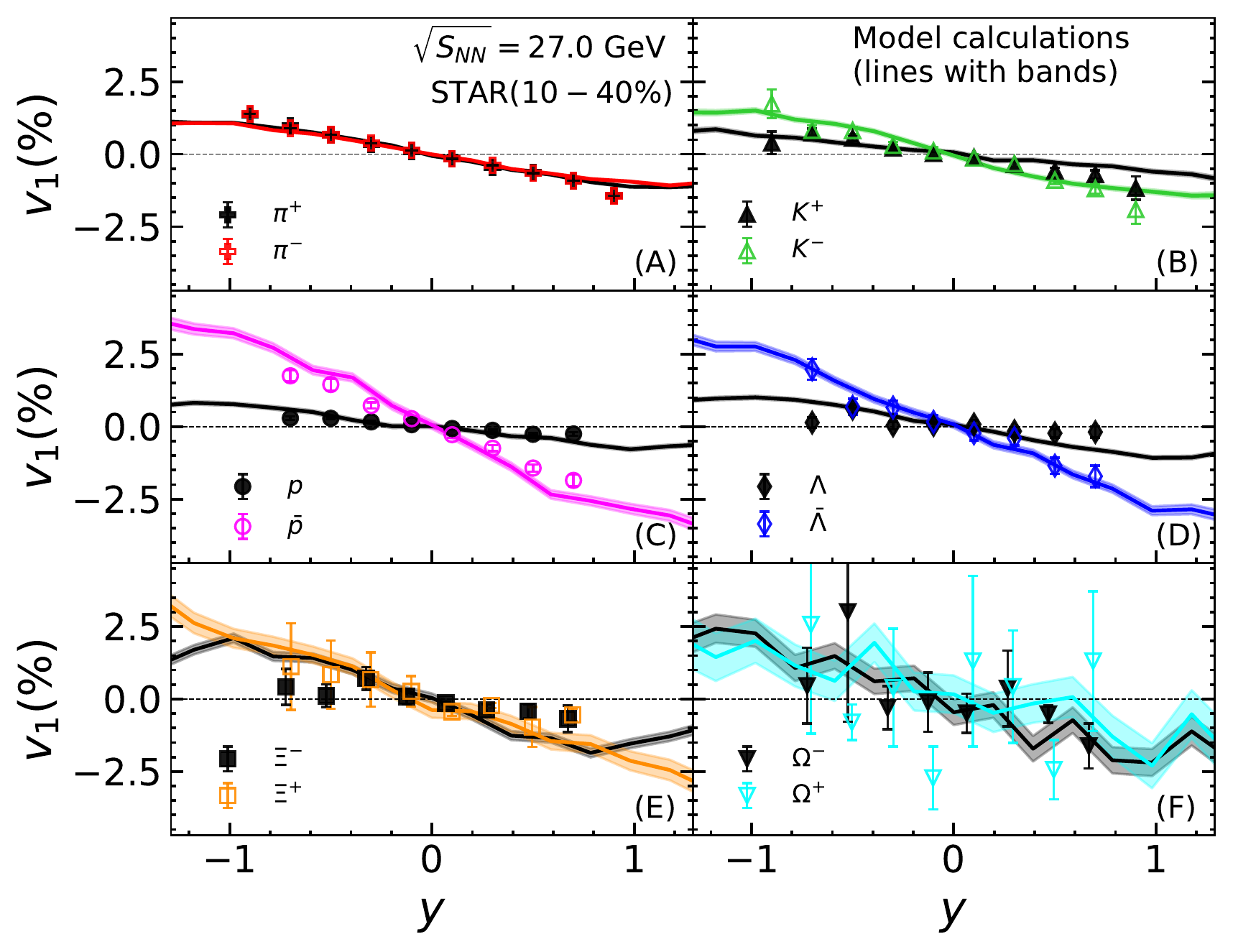}
 \caption{(Color online) The rapidity differential directed flow has been plotted for different identified hadrons in Au+Au collisions at $\sNN=27$ GeV. The model calculations with non-zero baryon diffusion are compared with the experimental measurements of STAR collaboration~\cite{STAR:2014clz,STAR:2017okv,STAR:2023wjl}. Directed flow of mesons are plotted in panel (A) and (B) where the black lines are for positively charged mesons and the colored lines are for negatively charged mesons. In panels (C), (D), (E) and (F), black lines show the model calculations of the baryons, while the colored lines are for anti-baryons.}
 \label{v1_y_all_identified_parts}
 \end{center}
\end{figure}

\begin{figure*}
 \begin{center}
  \includegraphics[scale=0.65]{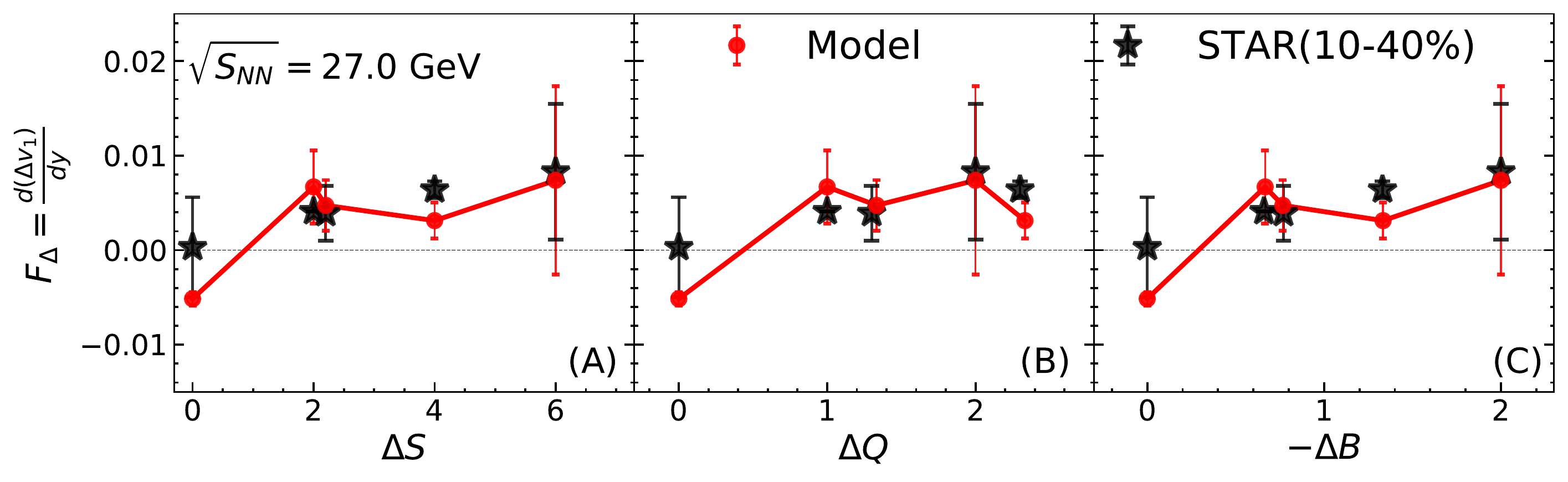}
 \caption{(Color online) The mid-rapidity slope of $\Delta v_1$ ($F_{\Delta}$) has been plotted as a function of $\Delta S$ (Panel-A), $\Delta Q$ (Panel-B) and $-\Delta B$ (Panel-C) for Au+Au
 collisions at $\sNN=27$ GeV. The model calculations with non-zero baryon diffusion are compared with the experimental measurements of STAR collaboration~\cite{STAR:2023wjl}. $F_{\Delta}$ is solely driven by the hydrodynamic response to baryon inhomogeneities in the model calculations.}
 \label{Dv1_Delta_only_27}
 \end{center}
\end{figure*}

In Fig.~\ref{v1_y_all_identified_parts}, we present the directed flow of identified hadrons plotted as a function of rapidity ($y$). The black lines show the model calculations of the baryons, while the colored lines are for anti-baryons. In the meson sector, the balck lines are for positively charged hadrons wheras the coloured lines are for negatively charged hadrons. The experimental measurements from the STAR collaboration~\cite{STAR:2014clz,STAR:2017okv,STAR:2023wjl} are shown in different symbols. The model calculations capture the $v_1(y)$ trend around the mid-rapidity region for all identified hadrons.

Another crucial observation is the split in $v_1$ between $K^{+}$ and $K^{-}$ due to the non-zero and inhomogeneous distribution of $\mu_S$ in the fireball. Even though we did not employ an independent initial distribution and evolution of net-strangeness, the split is obtained due to the imposed strangeness neutrality condition through EoS.

\begin{table}[ht]
\begin{tabular}{|p{0.9cm}|p{3.6cm}|p{0.9cm}|p{0.9cm}|p{0.9cm}|}
\hline  
Index & $\Delta v_1$ combinations & $\Delta Q$ & $\Delta S$ & $\Delta B$ \\
\hline  
1 & $\left[ \bar{p} + \phi \right] - \left[ K^{-} + \bar{\Lambda} \right]$ &  0 & 0 & 0 \\ 
2 & $\left[ \bar{\Lambda} \right] - \left[ \frac{1}{3} \Omega^{-} + \frac{2}{3} \bar{p} \right]$ & 1 & 2 & -2/3\\ 
3 & $\left[ \bar{\Lambda} \right] - \left[ K^{-} + \frac{1}{3} \bar{p} \right]$ & 4/3 & 2 &  -2/3\\ 
4 & $\left[ \bar{\Omega}^{+}  \right] - \left[  \Omega^{-} \right]$ & 2 & 6 & -2\\ 
5 & $\left[ \bar{\Xi}^{+} \right] - \left[ K^{-} + \frac{1}{3}  \Omega^{-}  \right]$ &  7/3 & 4 & -4/3\\ 
\hline
\end{tabular}
\caption{ Hadron combinations measured by the STAR collaboration~\cite{STAR:2023wjl}. }
\label{part_comb_table}
\end{table}

In a recent measurement by the STAR collaboration~\cite{STAR:2023wjl}, various identified particles are grouped in pairs where each pair has almost zero net mass difference at the
constituent quark level. The pairs are listed in Table~\ref{part_comb_table}. It is noteworthy that the net charge and net strangeness difference ($\Delta Q$ and $\Delta S$) between each pair is different, providing an opportunity to investigate $v_1$ splitting as a function of net charge or net strangeness difference by eliminating the background arising from the mass difference of constituent quarks~\cite{Sheikh:2021rew}. 
It has been observed that the $F_{\Delta} = d\Delta v_1/dy$ (slope of the difference of $v_1(y)$ between each considered pair) 
increases with $\Delta Q$, which was attributed to the effect of the EM field~\cite{STAR:2023wjl}.  

From our study, we have found that the considered pairs of combinations have a non-zero net baryon difference as well. 
So, we have calculated $F_{\Delta} $  in our model 
and plotted it as a function of $\Delta Q$, $\Delta S$ and $\Delta B$ in Fig.~\ref{Dv1_Delta_only_27}. The model calculation shows that $F_{\Delta} $ also increases with $\Delta B$ and is in quite good agreement with experimental measurements. 

This observation suggests that the dependency of $d\Delta v_1/dy$ on $\Delta Q$ might not be a clear signature of EM field. Although the pair combinations eliminate the background due to mass difference, the large background due to the physics of baryon stopping still persists.

Further, STAR collaboration has also measured the centrality dependence of the difference of $\frac{dv_1}{dy}$ 
between oppositely charged hadrons($\pi^{\pm}, K^{\pm}$ 
and $p(\bar{p})$), $\Delta \frac{dv_1}{dy}$~\cite{STAR:2023jdd}.
The observed negative $\Delta \frac{dv_1}{dy}$ at large centrality class is interpreted as the effect of electromagnetic field 
expected from the combined effect of Coulomb, Faraday, and Lorentz effects~\cite{Gursoy:2018yai} in Ref.~\cite{STAR:2023jdd}.
However, in our model calculations, we find negative $\Delta \frac{dv_1}{dy}$ at large centrality class from the physics of 
initial state baryon stopping alone without considering any EM field.

\begin{figure}
 \begin{center}
  \includegraphics[scale=0.65]{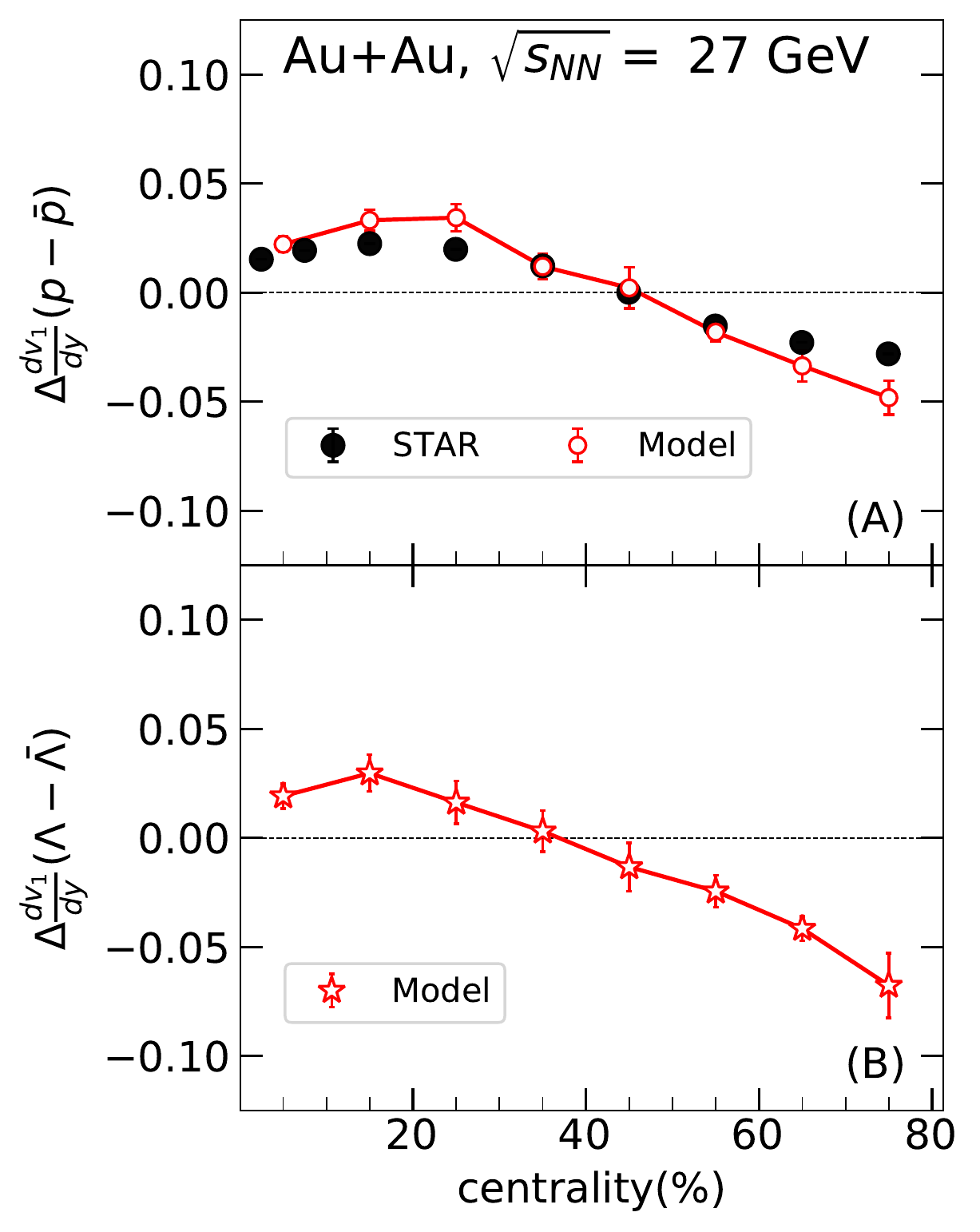}
 \caption{(Color online) The difference in the rapidity slope of baryons and anti-baryons as obtained in the hydrodynamic simulations 
 with non-zero baryon diffusion are compared to the experimental measurements of STAR collaboration~\cite{STAR:2023jdd}. Panel A shows 
 the observable for $\prp$ and Panel B shows the model prediction for $\Lambda$. The observable is solely driven by the hydrodynamic 
 response to baryon inhomogeneities in the model calculation.}
 \label{dv1dy_cent_baryon}
 \end{center}
\end{figure}

\begin{figure}
 \begin{center}
  \includegraphics[scale=0.65]{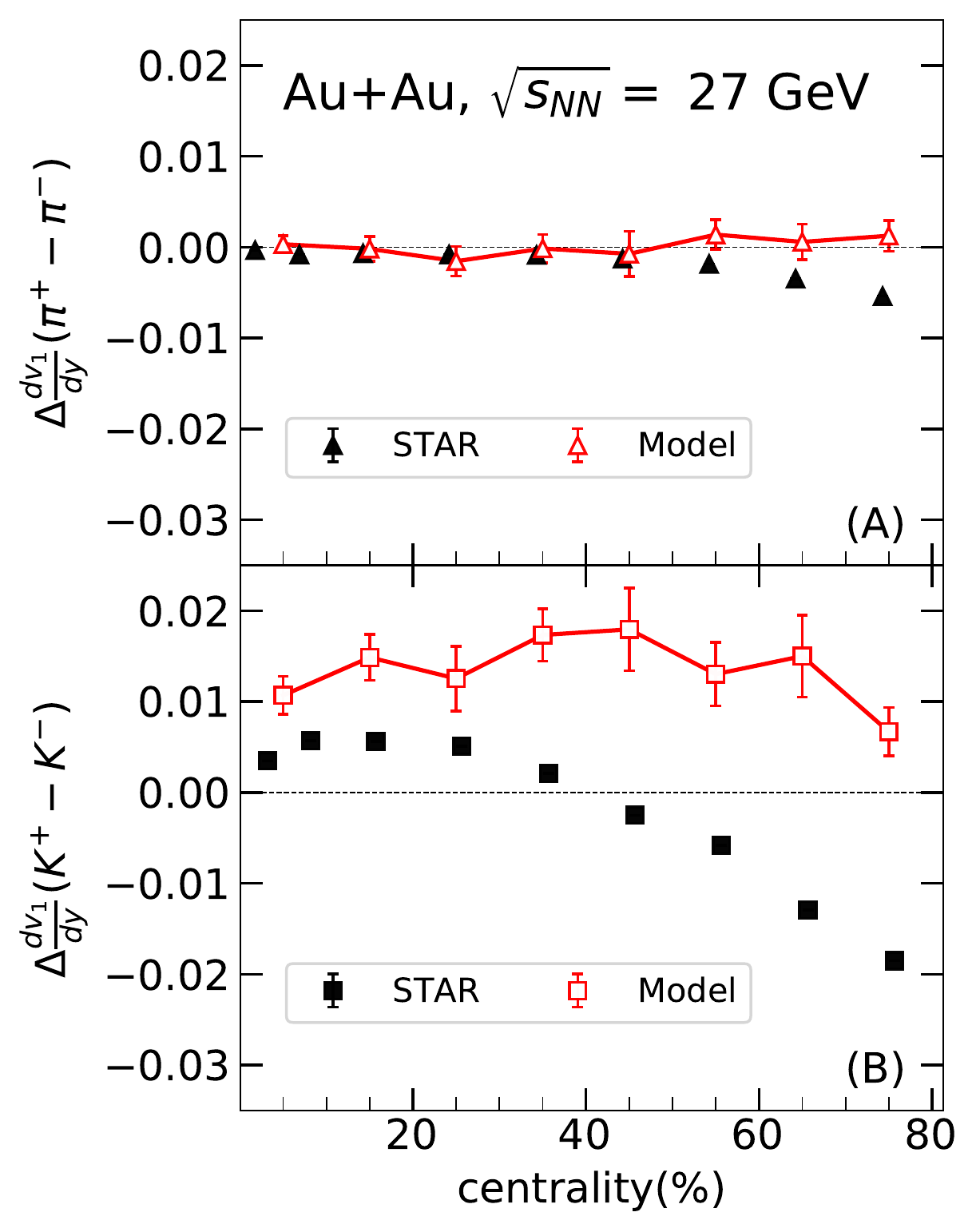}
 \caption{(Color online) The difference in the rapidity slope of oppositely charged mesons as obtained in the hydrodynamic simulations 
 with non-zero baryon diffusion are compared to the experimental measurements of STAR collaboration~\cite{STAR:2023jdd}. Panel A shows 
 the observable for $\pi$ and Panel B shows $\ka$. The observable is solely driven by the hydrodynamic response to baryon inhomogeneities 
 in the model calculation. The current calculation neglects electric charge and strangeness inhomogeneities that could affect this observable significantly for these hadrons.}
 \label{dv1dy_cent_meson}
 \end{center}
\end{figure}

The model calculation of the $\Delta \frac{dv_1}{dy}$ with centrality at Au+Au $\sNN=27$ GeV
has been plotted along with the recent measurements by STAR collaboration~\cite{STAR:2023jdd} in Figs.~\ref{dv1dy_cent_baryon} 
and \ref{dv1dy_cent_meson}. We first discuss the splitting of $\frac{dv_1}{dy}$ between protons and anti-protons, as shown in panel (A) of Fig.~\ref{dv1dy_cent_baryon}. The characteristic sign change from positive to negative values as one goes from more central to peripheral collisions that is observed by the STAR collaboration is also produced by the hydrodnamics simulations with baryon diffusion as a response to initial 
baryon inhomogeneities in the fireball. This underlines the strong contribution from baryon stopping physics to such signals of EM field. The model 
prediction for $\Lambda$ is shown in panel (B). The baryon inhomogeneities driven current also causes similar splitting and sign change 
versus centralities as in the case of protons. Being electric charge neutral, naively, one doesn't expect such a splitting for the lambdas due to 
the EM field. Hence, it will be very interesting for STAR to measure this observable for $\Lambda$.

Fig.~\ref{dv1dy_cent_meson} shows the results for the mesons. In this case, the current model results are unable to capture the $\Delta \frac{dv_1}{dy}$ between $\pi^{+}-\pi^{-}$ and $K^{+}-K^{-}$. This could be due to the fact that in our framework, we neither initialize nor evolve the strangeness and electric charge densities independently; rather, they are slaved to the baryon distribution by the relation $n_S=0$ and $n_{Q}=0.4 n_B$. However, this is not true in a realistic scenario, as locally inside the fireball, these conditions are violated. The independent inhomogeneous distribution and evolution of $n_S$ and $n_Q$ could lead to a different trend of $\Delta \frac{dv_1}{dy}(K^{+}-K^{-})$ or $\Delta \frac{dv_1}{dy}(\pi^{+}-\pi^{-})$ along centrality. Therefore, it is necessary to conduct a detailed study with a hydrodynamic framework consisting of all three conserved charge evolutions to better understand this phenomenon.

Based on our observations, we want to emphasize that the sign change in $\Delta \frac{dv_1}{dy}$ might not solely be attributed to the presence of a non-zero electromagnetic field. It receives a significant contribution from the conserved charges stopping and their subsequent evolution. Our calculations suggest that there may be a significant background contribution from conserved charge stopping physics, which makes it challenging to extract electromagnetic field signals from these observables. Therefore, a more comprehensive study is necessary to draw any firm conclusions at this time.

\section{summary}

Recently, STAR collaboration has presented two independent measurements which focus on the charge dependent directed flow splitting
in the presence of electromagnetic (EM) field. In one work, the difference of directed flow slope ($\Delta \frac{d v_1}{dy}$) between 
$\pi^{+}-\pi^{-}$, $K^{+}-K^{-}$ and $p-\bar{p}$  has been measured in different centralities. It has been observed 
that $\Delta \frac{d v_1}{dy}$ changes sign from positive in central collisions to negative in peripheral collisions. 
In another work, different particles are grouped in pairs where each pair has almost zero net mass difference at the constituent quark level.
However, the net charge difference ($\Delta Q$) in each pair of combinations are not same. STAR has observed that the the slope of the directed flow difference $F_{\Delta} = d\Delta v_1/dy, $ related to each combination pair increases with $\Delta Q$.
It is belived that both of the above discussed observations are the signature of EM field present in heavy ion collision. The results are interpreted as
a consequnce of Coulomb + Farday and Hall effect on the charged constituents. The objective of our study is to establish that the dependence of $\Delta \frac{d v_1}{dy}$ on centrality is not exclusively due to the effect of electromagnetic fields on charged constituents, but rather, 
there could be a significant contribution from the physics of conserved charge stopping. Additionally, 
we have demonstrated that the combination pairs used in the second analysis, which have a non-zero $\Delta Q$, 
also have a non-zero $\Delta B$. Therefore, the increase in $F_{\Delta}$ may not be solely due to the difference 
in charge, but rather, it is primarily a result of the difference in $\Delta B$.

We employed a multi-stage hybrid framework in our study, with a Glauber-based initial condition of energy and net-baryon distribution as input. This framework consists of viscous hydrodynamic evolution with baryon diffusion and late-stage hadronic interaction, and the model parameters were calibrated to capture the rapidity distribution of charged particle multiplicity, net-proton yield, and the rapidity-dependent directed flow of identified hadrons. Notably, we did not include any electromagnetic field contribution, yet we were able to describe the centrality dependence of directed flow splitting between proton and anti-proton. 

However, our model was unable to describe the $\Delta \frac{d v_1}{dy}$ between $\pi^{+}-\pi^{-}$ and $K^{+}-K^{-}$ which could be due to the lack of proper diffusion of net-strangeness and electric charge in our framework. We only considered the initial deposition and evolution of net-baryon, while the strangeness and electric charge densities were assumed to follow the relation, $n_S=0$ and $n_Q=0.4 n_B$. Hence, the interpretation of $\Delta \frac{d v_1}{dy}$ of mesons remains an open question. It will be interesting to consider the interplay of EM field and conserved charge dynamics on these observables.\\

\section{Acknowledgement}
The authors acknowledge helpful discussions with Sourendu Gupta, Sangyong Jeon, Subhash Singha 
and Nu Xu. SC acknowledges IISER Berhampur for Seed Grant.

\bibliographystyle{apsrev4-1}
\bibliography{manuscript}

\end{document}